# A Conceptual Approach to Two-Scale Constitutive Modelling For Hydro-Mechanical Coupling

G. D. Nguyen[1], A. El-Zein[2], and T. Bennett[3],

[1]Senior Lecturer, School of Civil, Environmental and Mining Engineering, University of Adelaide, Adelaide, SA 5005, Australia; PH +61 8 8313 2259; email: g.nguyen@adelaide.edu.au
[2]Associate Professor, School of Civil Engineering, University of Sydney, Sydney, NSW 2006, Australia; PH +61 2 9351 7351; email: abbas.elzein@sydney.edu.au
[3]Senior Lecturer, School of Civil, Environmental and Mining Engineering, University of Adelaide, Adelaide, SA 5005, Australia; PH +61 8 8313 6122; email: terry.bennett@adelaide.edu.au

**ABSTRACT**
Large scale modelling of fluid flow coupled with solid failure in geothermal reservoirs or hydrocarbon extraction from reservoir rocks usually involves behaviours at two scales: lower scale of the inelastic localization zone, and larger scale of the bulk continuum where elastic behaviour can be reasonably assumed. The hydraulic conductivities corresponding to the mechanical properties at these two scales are different. In the bulk elastic host rock, the hydraulic conductivity does not vary much with the deformation, while it significantly changes in the lower scale of the localization zone due to inelastic deformation. Increase of permeability due to fracture and/or dilation, or reduction of permeability due to material compaction can take place inside this zone. The challenge is to predict the evolution of hydraulic conductivities coupled with the mechanical behaviour of the material in all stages of the deformation process. In the early stage of diffuse deformation, the permeability of the material can be reasonably assumed to be homogenous over the whole Representative Volume Element (RVE) However, localized failure results in distinctly different conductivities in different parts of the RVE. This paper establishes a general framework and corresponding field equations to describe the hydro-mechanical coupling in both diffuse and localized stages of deformation in rocks. In particular, embedding the lower scale hydro-mechanical behaviour of the localization zone inside an elastic bulk, together with their corresponding effective sizes, helps effectively deal with scaling issues in large-scale modelling. Preliminary results are presented which demonstrate the promising features of this new approach.

## 1 INTRODUCTION

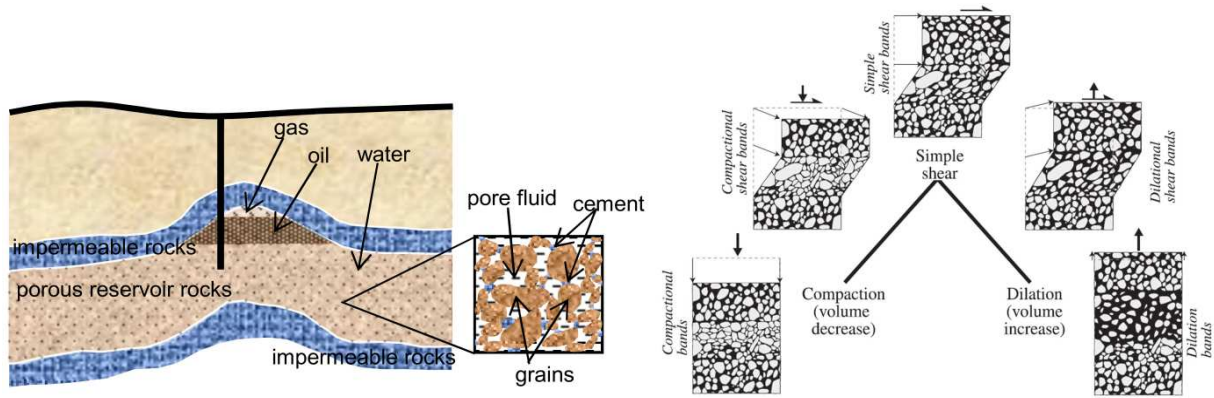

(a): Oil extraction and porous reservoir rock    (b): Localised failure modes (Fossen et al, 2007)
Figure 1: Porous reservoir rock and its localised failure modes.

It is well known that localised deformation in high porosity rocks can significantly affect the permeability of the host rock (Caine et al, 1996; Evans et al, 1997; Tenthorey et al, 2003; Zhu and Wong, 1997) and hence the flow of hydrocarbon during extraction. Grain scale mechanisms such as grain crushing, cement debonding and pore collapse are identified as the sources of this localised failure and the integration of these mechanisms is essential in modelling and simulation. The issues with large scale modelling, as identified in the abstract, is the connection between what happens at the grain scale where inelastic deformation takes place and the macro scale of the continuum modelling. In particular, the localisation band only occupies a very small volume of the modelled domain, and discretisation to this size (width of localisation band, or the orders of the grain sizes) is impossible given the spatial size of the problems (several kilometres) and available computing resource. In terms of hydro-mechanical coupling, the localisation band can act as a fluid flow barrier due to grain crushing and compaction, or fluid pathway due to shear dilation. It is therefore essential that these grain scale phenomena and their effects on the macro hydro-mechanical modelling be properly captured in the simulation.

A classical constitutive modelling approach is unable to capture localised phenomena at sub-RVE scale, due to the assumption of homogeneity of deformation over the whole representative volume element (RVE). The state variables and material properties are assumed to be uniformly distributed over the whole RVE. As seen in Fig. 1b, this homogeneity assumption is no longer valid once localised deformation occurs, resulting in the invalidity of the models in localised stage of failure. In modelling the mechanical behaviour, the issues of localised failure and size effects are well known and enrichments of the RVE are needed to properly capture the localised stage of failure that usually follows homogeneous stage (see Fig. 2). Simple treatment to scale/smear the inelastic behaviour inside the localisation zone with the resolution of the spatial discretisation, e.g. the smeared crack/deformation approach (Bazant and Oh, 1983; Crook et al, 2006), can work for small problems, but encounters unphysical constitutive snap back in large scale problems. In terms of large scale hydro-mechanical modelling, enrichments to the constitutive structure are therefore needed to accommodate the size of the localisation and both its hydraulic and mechanical properties. In this context, models based on the dual-porosity approach (Warren and Root, 1963; Kazemi, 1969) are unable to address the issues of localised mechanical failure coupled with hydraulic conductivity, as this approach is built on classical continuum mechanics with the assumption of homogeneous deformation over the whole RVE and regularly repetitive pore structures over the domain.

To deal with issues related to localised failure and size effects, the enrichment to constitutive modelling structure (Nguyen et al, 2012 & 2014; Nguyen, 2014) has been shown to have advantages over element based enrichment techniques. This is because it is able to deal with complex localisation (or fracture) patterns that are hard or even impossible for other element based enrichments such as the enhanced strain approach (e.g. Larsson et al, 1996; Oliver, 1996; Borja, 2000; Wells & Sluys, 2001a; Foster et al., 2007) or extended finite element method (e.g. Wells and Sluys, 2001b; Samaniego & Belytschko, 2005; Borja, 2008; Sanborn & Prevost, 2011). In the context of hydro-mechanical coupling, the work by Rethore et al (2007) can be referred to, as an example of an element based enrichment for tackling both mechanical and fluid flow behaviour. In this work the approach proposed in Nguyen et al (2012 & 2014) for enhancing the constitutive structure is adopted, as it provides a good basis for tackling large scale issues. In addition it does not require the reformulation of finite elements to incorporate non-homogeneous kinematic fields that occur due to localisation, while being able to capture localised failure and size effects. As a result, the implementation in any existing numerical codes is straightforward as everything is contained in the constitutive routine.

The paper is organised as follows. Background in enriching a model with the behaviour associated with the localisation band is described in section 2.1. The mechanical enrichment will open a way to introduce terms related to the hydraulic conductivity and the hydro-mechanical couplings. Different scenarios associated with the hydraulic and mechanical behaviour of the localisation band are discussed and corresponding treatments addressed.

## 2 TWO-SCALE CONSTITUTIVE MODELLING
### 2.1 Enhancements to the mechanical behaviour
The focus of the enrichment is the post peak behaviour of the material, when strain softening leads to the localisation of inelastic behaviour on a narrow band, while outside this band the material is elastically unloading (Fig. 2). In such a case the deformation over a RVE on which a constitutive model is defined is strongly inhomogeneous (Figs. 1b & 2) and classical approaches with the assumptions of homogeneity are no longer valid. As a consequence, this localised failure mode cannot be properly captured using classical plasticity-type models, which are only valid for the pre-peak regime with either elastic or inelastic homogenous deformation over the whole RVE. In this paper, we treat the RVE as a composite material with two separate behaviours for the volume inside and outside the localisation band. In continuum modelling, this allows keeping an invariant size of the localisation zone embedded in an elastic media, as the size of the localisation zone is considered as a physical property associated with a material and loading conditions. Later we will see that this open a new way to couple the material behaviour with its hydraulic conductivity via the effects of mechanical deformation in the localisation band on the permeability of the material.

For a volume element with an embedded localisation band treated as a composite material, the following volume fraction is defined as (see Fig. 1):

$$\eta = \frac{\Omega_i}{\Omega} = \frac{h}{H} \tag{1}$$

where $h$ and $H$ are correspondingly the width of the localisation band and effective size of the volume element. The macro strain rate in the solid skeleton (denoted by superscript $s$) can then be defined using volume averaging taking into account separate strains inside and outside the localisation band:

$$\dot{\varepsilon}^s = \eta \dot{\varepsilon}_i^{\,s} + (1-\eta)\dot{\varepsilon}_o^{\,s} \tag{2}$$

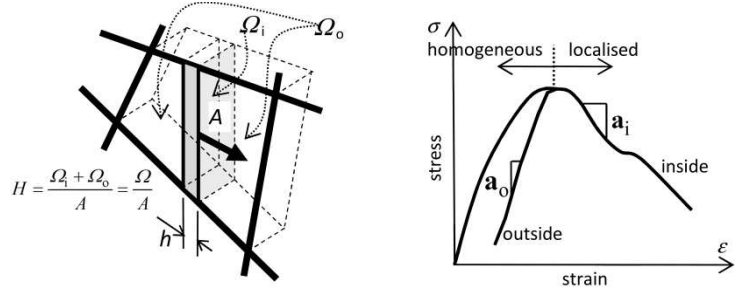

Figure 2: Numerical discretisation and a localization zone (darkened) (after Nguyen et al, 2012)

Given the assumption that the size $h$ of the band is usually very small compared to the effective size of the volume element, the strain inside the localization zone can be expressed in terms of the relative displacement [**u**] between the two sides of the localisation band (Nguyen et al, 2012 & 2014). In incremental form, it reads

$$\dot{\varepsilon}_i^{\,s} = \tfrac{1}{h}\left(\mathbf{n}\otimes[\dot{\mathbf{u}}]\right)^{sym} = \tfrac{1}{2h}\left(\mathbf{n}\otimes[\dot{\mathbf{u}}]+[\dot{\mathbf{u}}]\otimes\mathbf{n}\right) \tag{3}$$

The constitutive behaviours of the materials inside and outside the localisation band are expressed through the corresponding stress-strain relationships $\sigma_i$-$\varepsilon_i$ and $\sigma_o$-$\varepsilon_o$, respectively. While the deformation across the RVE is inhomogeneous, those inside and outside the localisation band (Figs. 1b & 2) can be reasonably considered as homogeneous and also isotropic. Since the differential equilibrium equation is written for the whole composite volume element using the macroscopic stress $\boldsymbol{\sigma}$, virtual work equation is needed to link this macroscopic stress with the lower scale stresses inside and outside the localisation band. For this, we can write:

$$\boldsymbol{\sigma}:\dot{\boldsymbol{\varepsilon}}^s = \eta\boldsymbol{\sigma}_i:\dot{\boldsymbol{\varepsilon}}_i^{\,s} + (1-\eta)\boldsymbol{\sigma}_o:\dot{\boldsymbol{\varepsilon}}_o^{\,s} \tag{4}$$

Substituting (2) and (3) into (4) and rearranging the obtained expression leads to

$$\tfrac{\eta}{h}(\mathbf{t}-\mathbf{t}_i)\cdot[\dot{\mathbf{u}}] + (1-\eta)(\boldsymbol{\sigma}-\boldsymbol{\sigma}_o):\dot{\boldsymbol{\varepsilon}}_o^{\,s} = 0 \tag{5}$$

where $\mathbf{t}=\boldsymbol{\sigma}\cdot\mathbf{n}$ and $\mathbf{t}_i=\boldsymbol{\sigma}_i\cdot\mathbf{n}$ are the tractions at the interface of the localisation band. The above condition must be met for any $[\dot{\mathbf{u}}]$ and $\dot{\boldsymbol{\varepsilon}}_o^{\,s}$, resulting in the following requirements:

$$\mathbf{t}=\mathbf{t}_i \text{ and } \boldsymbol{\sigma}=\boldsymbol{\sigma}_o \tag{6}$$

It can be seen that the above two conditions lead to the traction continuity across the interface of the localisation band, $\mathbf{t}_o=\mathbf{t}_i$. Since $\boldsymbol{\sigma}=\boldsymbol{\sigma}_o$, the structure of the constitutive equations then becomes:

Macro scale constitutive relationship: $\dot{\boldsymbol{\sigma}} = \dot{\boldsymbol{\sigma}}_o = \mathbf{a}_o:\dot{\boldsymbol{\varepsilon}}_o^{\,s} = \tfrac{1}{1-\eta}\mathbf{a}_o:\left(\dot{\boldsymbol{\varepsilon}}^s - \eta\dot{\boldsymbol{\varepsilon}}_i^{\,s}\right)$ (7)

Constitutive relationship for the localization zone: $\dot{\boldsymbol{\sigma}}_i = \mathbf{a}_i:\dot{\boldsymbol{\varepsilon}}_i^{\,s} = \mathbf{a}_i:\tfrac{1}{h}\left(\mathbf{n}\otimes[\dot{\mathbf{u}}]\right)^{sym}$ (8)

Internal equilibrium condition connecting macro and lower scale behaviour: $(\dot{\boldsymbol{\sigma}}_o - \dot{\boldsymbol{\sigma}}_i)\cdot\mathbf{n} = 0$ (9)

The above equations are written in generic form, using tangent stiffnesses $\mathbf{a}_i$ and $\mathbf{a}_o$ associated with the constitutive behaviours inside and outside the localisation band. From the above 3 equations, the macro stress-strain rate relationships can be worked out (Nguyen et al, 2012):

$$\dot{\boldsymbol{\sigma}} = \tfrac{1}{1-\eta}\mathbf{a}_o:\left\{\dot{\boldsymbol{\varepsilon}}^s - \tfrac{1}{H}\left[\mathbf{n}\otimes\left(\mathbf{C}^{-1}\cdot(\mathbf{a}_o:\dot{\boldsymbol{\varepsilon}}^s)\cdot\mathbf{n}\right)\right]^{sym}\right\} \tag{10}$$

where the lower scale behaviour is embedded in **C** that combines both sizes (H and h) and behaviours ($\mathbf{a}_o$ and $\mathbf{a}_i$) associated with macro and lower scales:

$$\mathbf{C} = \tfrac{1}{H}\mathbf{n}\cdot\mathbf{a}_o\cdot\mathbf{n} + (1-\eta)\tfrac{1}{h}\mathbf{n}\cdot\mathbf{a}_i\cdot\mathbf{n} \tag{11}$$

The localization is triggered by the loss of positiveness of the acoustic tensor in homogeneous mode preceding this localized stage. Due to the involvement of an oriented band, the overall constitutive behaviour is anisotropic regardless of the isotropy of the materials inside and outside the localisation band. This mechanical feature has been demonstrated in Nguyen et al (2014).

## 2.2 Enhancements to the hydraulic conductivity

Generally, localised failure leads to inelastic deformation associated with changes in the material microstructures inside the localised region while outside this region the material behaviour can be reasonably assumed to be elastic with negligible microstructural change. The localisation band in this case can act as a fluid pathway or flow barrier depending on the materials and loading conditions. As examples, hydraulic fracturing of hard rocks opens fluid channels accelerating the flows, while in cemented granular rocks, localised deformation can be associated with either compaction or dilation of the material. In particular, at low confining pressure, shear deformation usually leads to dilation of the material, with breaking of cement bridging the rock grains followed by sliding of the grains. On the other hand, shearing these porous rocks under high confining pressure will crush the grains and cements, which consequently triggers the collapse of pores. In the former case of dilation band, increase of permeability is expected while compaction due to grain crushing in the latter case results in the reduction of the permeability, hence decelerating the flow velocity across the localisation band.

In line with the above generic descriptions of mechanical behaviour, homogeneity assumption on the permeabilities of the materials inside and outside the localisation band can be reasonably made. In such a case, the permeability tensors $\mathbf{k}_i$ and $\mathbf{k}_o$ are both isotropic and can be expressed in two-dimensional (2D) cases as:

$$\mathbf{k}_i = \begin{bmatrix} k_i & 0 \\ 0 & k_i \end{bmatrix} \text{ and } \mathbf{k}_o = \begin{bmatrix} k_o & 0 \\ 0 & k_o \end{bmatrix} \quad (12)$$

Similar to the case of momentum equation for the mechanical behaviour written in terms of the macroscopic stress $\boldsymbol{\sigma}$, we want to write the mass conservation equation for the whole composite volume with an embedded localisation band. Assuming no sink source, the mass conservation can be written as (Chen et al, 2006):

$$\frac{\partial(\phi\rho)}{\partial t} = -\nabla \cdot [\rho(\mathbf{v}^f - \mathbf{v}^s)] = -\nabla \cdot (\rho\mathbf{v}) \quad (13)$$

where $\mathbf{v}^f$ is fluid velocity, $\mathbf{v}^s$ solid velocity, $\rho$ fluid density per unit volume, and $\phi$ solid porosity. Ignoring gravity, Darcy law governing the fluid flow can be written as:

$$\mathbf{v} = \mathbf{v}^f - \mathbf{v}^s = -\frac{\mathbf{k}}{\mu}\nabla p \quad (14)$$

The relative velocity $\mathbf{v}$ between the fluid and solid is governed by the pressure gradient $\nabla p$, permeability $\mathbf{k}$ of the volume element and fluid viscosity $\mu$. Since localised failure leads to the microstructural changes of the material inside the localisation zone, the permeability tensor $\mathbf{k}$ of the whole volume is governed by both $\mathbf{k}_i$ and $\mathbf{k}_o$ and their corresponding volume fractions. While $\mathbf{k}_o$ is relatively invariant during elastic deformation outside the localisation band, $\mathbf{k}_i$ is evolving with the microstructural changes in this band. In the context of modelling crushable granular materials, the readers are referred to Nguyen & Einav (2009) for an explicit link between permeability and inelastic mechanical behaviour via the evolving grain size distribution.

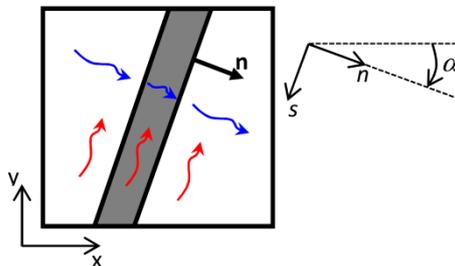

Figure 3: Flow through a composite volume.

Given the flow across a volume crossed by a localisation band in Fig. 3, it can be seen that the flow rate perpendicular to the band is controlled by both the permeability of the band and that of the surrounding area. In this work, we treat the flows as two independent 1D problems for the $n$ (normal) and $s$ (tangent) directions in the local coordinate system $ns$ (Fig. 3b). In the general case with $\mathcal{H}$ being the total hydraulic head, we apply a constant hydraulic gradient $\partial \mathcal{H}/\partial n$ across the volume with a localisation band and find the equivalent hydraulic conductivity analytically by forcing continuity of the hydraulic head and hydraulic flux (Darcy velocity) at the interface between the localisation band and the outside region. Defining an equivalent hydraulic conductivity for the volume with a localisation band as (i.e. homogenizing Darcy's law over this volume):

$$k_n = -\frac{\frac{1}{A}\int_A v_n dA}{\partial \mathcal{H}/\partial n} \quad (15)$$

where $v_n$ is the component in the $n$ direction of the Darcy velocity. It can be shown that (see Appendix for details):

$$k_n = \frac{k_o k_i}{(1-\eta)k_i + \eta k_o} \quad (16)$$

Doing the same in the $s$ direction, with a constant $\partial \mathcal{H}/\partial s$ applied across the volume with a localisation band, we obtain:

$$k_s = -\frac{\frac{1}{A}\int_A v_s dA}{\partial \mathcal{H}/\partial s} \quad (17)$$

and

$$k_s = \eta k_i + (1-\eta)k_o \quad (18)$$

Hence, denoting **k'** the permeability tensor in the local coordinate system *n-s* attached to the localisation band (Fig. 3), we can write:

$$\mathbf{k}' = \begin{bmatrix} k_n & 0 \\ 0 & k_s \end{bmatrix} = \begin{bmatrix} \frac{k_o k_i}{(1-\eta)k_i + \eta k_o} & 0 \\ 0 & \eta k_i + (1-\eta)k_o \end{bmatrix} \quad (19)$$

In the *x-y* Cartesian coordinate system, we will have the permeability tensor **k** in the form:

$$\mathbf{k} = \mathbf{R}^T \mathbf{k}' \mathbf{R} \quad (20)$$

where **R** is the rotation matrix determined from the orientation of the band (Fig. 3):

$$\mathbf{R} = \begin{bmatrix} n_x & -n_y \\ n_y & n_x \end{bmatrix} = \begin{bmatrix} \cos\alpha & -\sin\alpha \\ \sin\alpha & \cos\alpha \end{bmatrix} \quad (21)$$

Therefore we can write:

$$\mathbf{k} = \begin{bmatrix} n_x^2 k_n + n_y^2 k_s & n_x n_y(-k_n + k_s) \\ n_x n_y(-k_n + k_s) & n_y^2 k_n + n_x^2 k_s \end{bmatrix} \quad (22)$$

### 2.3 Coupled equations

Neglecting gravity and assuming fluid incompressibility, the coupled hydro-mechanical system of equations can be written in general form as:

Momentum conservation: $\nabla \cdot (\boldsymbol{\sigma} + p\mathbf{I}) = 0$ (23)

Mass conservation: $\frac{\partial(\phi\rho)}{\partial t} = -\nabla \cdot [\rho(\mathbf{v}^f - \mathbf{v}^s)]$ (24)

Constitutive equations for solid: $\dot{\boldsymbol{\varepsilon}}^s = (\nabla \mathbf{v}^s)^{sym} = \frac{1}{2}\left(\frac{\partial v_i^s}{\partial x_j} + \frac{\partial v_j^s}{\partial x_i}\right)$, and $\dot{\boldsymbol{\sigma}} = \mathbf{D}^T : \dot{\boldsymbol{\varepsilon}}^s$ (25)

Constitutive equations for fluid: $\dot{\boldsymbol{\varepsilon}}^f = (\nabla \mathbf{v}^f)^{sym} = \frac{1}{2}\left(\frac{\partial v_i^f}{\partial x_j} + \frac{\partial v_j^f}{\partial x_i}\right)$, and $\mathbf{v}^f - \mathbf{v}^s = -\frac{\mathbf{k}}{\mu}\nabla p$ (26)

In the above system, a general form of mechanical constitutive behaviour is used, and the tangent stiffness $\mathbf{D}^T$ is derived from equation (10) that links the macroscopic stress rate with macroscopic strain rate through the behaviours inside and outside the embedded localisation band, the corresponding sizes and orientation of the band.

## 3 DISCUSSION

The above system of equations (23-26) governs the coupled hydro-mechanical behaviour and needs to be solved numerically. While this is a step further from the current proposed modelling approach, a discussion of several aspects of the modelling and analysis is provided in this section, with focus on the localised stage of failure and the hydro-mechanical coupling within it.

### 3.1 Diffuse and localised failure

The system of equations (23-26) describes the hydro-mechanical coupling during diffuse and localised stages of failure in a general form. The diffuse (or homogeneous) stage precedes the localised one and can be considered as a special case with the same constitutive behaviour inside and outside the localisation band. Mathematically it is described by the constitutive tangent stiffness, $\mathbf{a}_i = \mathbf{a}_o$, and isotropic permeability, $\mathbf{k}_i = \mathbf{k}_o$, for the whole volume element. Localised stage of failure is triggered by the loss of positiveness of the acoustic tensor $\mathbf{A}$:

$$\det(\mathbf{A}) = \det(\mathbf{n} \cdot \mathbf{a}_i \cdot \mathbf{n}) \leq 0. \tag{27}$$

Therefore during the homogeneous stage, this condition is checked at every time step for every possible orientation $\mathbf{n}$. The onset of localised failure and corresponding orientation of the localisation band is determined from the following condition (Das et al, 2013):

$$\min_\mathbf{n} \det(\mathbf{A}) = \min_\mathbf{n}[\det(\mathbf{n} \cdot \mathbf{a}_i \cdot \mathbf{n})] \leq 0. \tag{28}$$

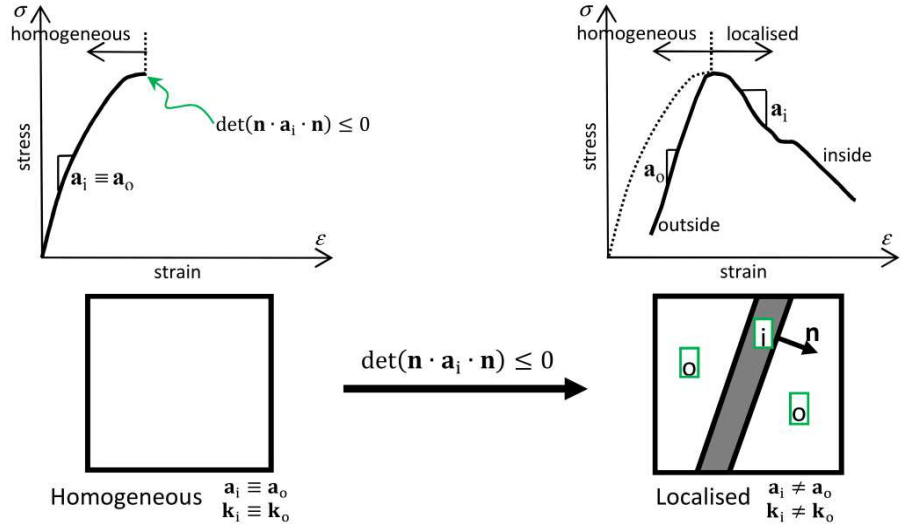

Figure 4: Hydro-mechanical coupling during diffuse and localised stages.

At the onset of localisation, a localisation band with bandwidth $h$ governed by the microstructure (e.g. grain size distribution) is inserted into the constitutive description. The mechanical behaviour then follows different branches with different tangent stiffness, $\mathbf{a}_i \neq \mathbf{a}_o$ (Fig. 4). As a consequence, the permeabilities of the material inside and outside the localisation are evolving with the inelastic microstructural changes and elastic unloading, respectively (Fig. 4). The mechanical part is described by the enriched constitutive relationship (10), while its effects on the overall permeability of the volume element is characterised by equation (22). Despite the isotropy of permeability tensors, the composite one characterising the hydraulic conductivity of the whole RVE is anisotropic due to different flow rate inside and outside the localisation region in conjunction with the oriented band.

### 3.2 Special cases of localisation band
*Dilation band or fracture*, with $k_i \gg k_o$: The permeabilities in the normal and tangent directions (with respect to the band) are, respectively:

$$k_n = \frac{k_o k_i}{\eta k_o + (1-\eta) k_i} = \frac{k_o}{\eta k_o/k_i + (1-\eta)} \approx \frac{k_o}{1-\eta} \tag{29}$$

$$k_s = \eta k_i + (1-\eta) k_o \approx \eta k_i \tag{30}$$

While the flow in the tangent direction is mainly controlled, or accelerated, by the fracture, that in the normal direction is still governed by the permeability of the host rock.

*Compaction band*, with $k_i \ll k_o$: The permeabilities in the normal and tangent directions (with respect to the band) are, respectively:

$$k_n = \frac{k_o k_i}{\eta k_o + (1-\eta) k_i} = \frac{k_i}{\eta + (1-\eta) k_i/k_o} \approx \frac{k_i}{\eta} \tag{31}$$

$$k_s = \eta k_i + (1-\eta) k_o \approx (1-\eta) k_o \tag{32}$$

In this case, compaction band acts as a flow barrier blocking the flow in the normal direction, while the tangent flow is mainly controlled by the permeability of the host rock.

### 3.3 Size effects
For the proposed coupling strategy, both mechanical and hydraulic behaviours scale with size $h$ of the localisation band and the effective size $H$ of the composite volume crossed by this localisation band. While $h$ is a material property and hence invariant with the numerical scheme, $H$ depends on the discretisation and in the numerical implementation is associated with the size of a finite element (Nguyen et al, 2014). As a consequence it varies with the discretisation. Details on how to determine this effective size $H$ along with implementation algorithms for the mechanical behaviour are described at length in Nguyen et al (2014). The key point of the proposed modelling strategy is the use of a physical size associated with the material in the constitutive descriptions, instead of only scaling the behaviour with the size *H* of the discretisation, as seen in the classical approach.

In the context of finite element modelling, the volume occupied by an integration point varies with the resolution of the mesh, and it is implicitly assumed in this proposed approach that the localisation band crosses the whole volume of the RVE, i.e. the integration point in FEM. The limitation hence lies in the resolution of the discretisation, as it should be sufficiently fine to capture the extension of the localisation band. Preliminary results on the mechanical behaviour (Nguyen et al, 2014) shows that this is not a serious issue, and provided the resolution is sufficient to capture the propagation, the overall behaviour and localisation pattern are independent of the discretisation. Since the extension *L* of the band is several orders of magnitude higher than its width (e.g. *L>>h*), the issues of classical approach related to the spatial discretisation and large scale modelling mentioned in the introduction can be overcome. This is because the width *h* of the band has already been embedded in the constitutive description, and hence the resolution does not need to be of the same order of magnitude as the width of the localisation band to properly capture its mechanical and physical behaviour.

## 4 CONCLUSIONS

We proposed a conceptual approach to hydro-mechanical coupling taking into account the localised nature of failure in geomaterials. The generic approach is developed for a volume with an embedded localisation band which can act either as a compaction band with reduced permeability, or as a shear dilation band with increased permeability. The orientation of the band, once introduced, naturally results in anisotropic hydro-mechanical behaviour. This is an important aspect in modelling flow channelling or flow barrier, besides the ability to capture size effects thanks to the possession of an intrinsic length scale in the constitutive structure. The next stage in this research project is the implementation of the formulation and its application to specific problems.

## 5 ACKNOWLEDGEMENTS

Giang Nguyen and Abbas El-Zein acknowledge funding support from the Australian Research Council via project DP140100945.## 6 APPENDIX

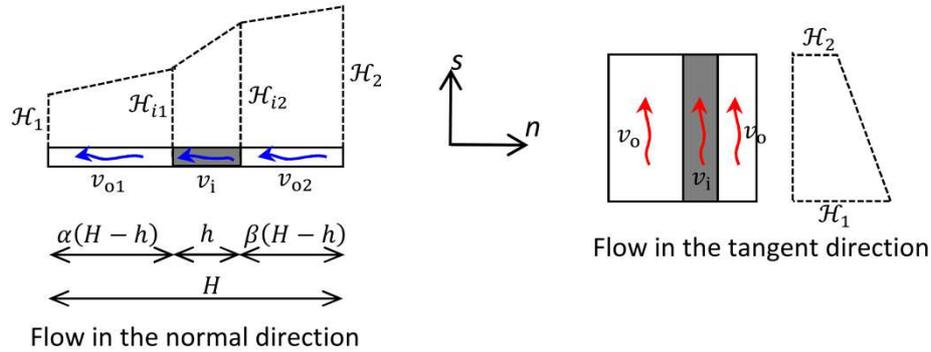

Figure A1: Decomposition of the flow into two 1D problems.

For the direction normal to the band, by definition the overall permeability of the composite volume is related to the velocity $v_n$ by:

$$v_n = -\frac{k_n}{\mu}\frac{\mathcal{H}_2-\mathcal{H}_1}{H} \quad (A1)$$

In the same sense, the fluid velocity in different regions of the composite volume (Fig. A1) can be written as:

$$v_{o1} = -\frac{k_o}{\mu}\frac{\mathcal{H}_{i1}-\mathcal{H}_1}{\alpha(H-h)}; \quad v_{o2} = -\frac{k_o}{\mu}\frac{\mathcal{H}_2-\mathcal{H}_{i2}}{\beta(H-h)}; \quad v_i = -\frac{k_i}{\mu}\frac{\mathcal{H}_{i2}-\mathcal{H}_{i1}}{h}, \text{ where } \alpha+\beta=1 \quad (A2)$$

Enforcing continuity of fluid velocity across the interfaces of the localisation band leads to:

$$\mathcal{H}_1 = \mathcal{H}_{i1} - \frac{\alpha(H-h)}{h}\frac{k_i}{k_o}(\mathcal{H}_{i2}-\mathcal{H}_{i1}) \quad (A3)$$

$$\mathcal{H}_2 = \mathcal{H}_{i2} + \frac{\beta(H-h)}{h}\frac{k_i}{k_o}(\mathcal{H}_{i2}-\mathcal{H}_{i1}) \quad (A4)$$

Therefore using the volume fraction $\eta = h/H$ (see equation 1), we obtain:

$$\mathcal{H}_2 - \mathcal{H}_1 = \left(1+\frac{H-h}{h}\frac{k_i}{k_o}\right)(\mathcal{H}_{i2}-\mathcal{H}_{i1}) = \frac{\eta k_o + (1-\eta)k_i}{\eta k_o}(\mathcal{H}_{i2}-\mathcal{H}_{i1}) \quad (A5)$$

Since the fluid velocity is uniform across the length, the permeability can then be determined from the condition $v_n = v_i$:

$$-\frac{k_n}{\mu}\frac{\mathcal{H}_2-\mathcal{H}_1}{H} = -\frac{k_i}{\mu}\frac{\mathcal{H}_{i2}-\mathcal{H}_{i1}}{h} \tag{A6}$$

Substituting (A5) into (A6) and rearranging the obtained expression, we get:

$$k_n = \frac{k_o k_i}{\eta k_o + (1-\eta)k_i} \tag{A7}$$

For the flow in the tangent direction, as seen in Fig. A1, the continuity across the interfaces of the localisation band is not required, resulting in a volume averaging of the permeability, given the same hydraulic gradient $\partial \mathcal{H}/\partial s$ applied to both regions.